\documentclass[aps,onecolumn,letterpaper,oneside,preprint,tightenlines,draft,%
nobibnotes,nofootinbib,amsfonts,amssymb,amsmath,showpacs,showkeys]{revtex4}
\newcommand{\cN}{\mathcal{N}}
\newcommand{\bbR}{\mathbb{R}}

\newcommand{\rmd}{\mathrm{d}}%neIOPART

\begin{document}

%\markboth{Authors' Names}
%{Instructions for Typing Manuscripts (Paper's Title)}%WSPC

%\catchline{}{}{}{}{}%WSPC
%
%%%%%%%%%%%%%%%%%%%%%%%%%%%%%%%%%%%%%%%%%%%%%%%%%%%%%%%%%%%%%%%%%%%%

\title{$\cN$-fold Supersymmetry in Quantum Mechanical Matrix Models}
%\article[Short title]{TYPE}{Full title}%IOPART
%ELSART,WSPC
%\author{},
%\ead{}%ELSART
%\address{}
%\author{Toshiaki Tanaka}
%\ead{ttanaka@mail.ncku.edu.tw}%ELSART
%\address{Department of Physics, National Cheng Kung University,\\
% Tainan 701, Taiwan, R.O.C.}
%\address{National Center for Theoretical Sciences, Taiwan, R.O.C.}
%REVTEX4
\author{Toshiaki Tanaka}
\email{toshiaki@post.kek.jp}
\affiliation{Institute of Particle and Nuclear Studies,
 High Energy Accelerator Research Organization (KEK),
 1-1 Oho, Tsukuba, Ibaraki 305-0801, Japan}
%  \altaffiliation{}
%IOPART
%\author{Author 1\dag, Author 2\ddag and Toshiaki Tanaka\S}
%\address{\dag Address 1}
%\address{\ddag Address 2}
%\address{\S Institute of Particle and Nuclear Studies,
% High Energy Accelerator Research Organization (KEK),
% 1-1 Oho, Tsukuba, Ibaraki 305-0801, Japan}
%\ead{toshiaki@post.kek.jp}
%\eads{\mailto{email 1}, \mailto{email 2},
% \mailto{toshiaki@post.kek.jp}}

%\date{\today}

\begin{abstract}

We formulate $\cN$-fold supersymmetry in quantum mechanical matrix models.
As an example, we construct general two-by-two Hermitian matrix $2$-fold
supersymmetric quantum mechanical systems. We find that there are two
inequivalent such systems, both of which are characterized by two arbitrary
scalar functions, and one of which does not reduce to the scalar system.
The obtained systems are all weakly quasi-solvable.

%\keywords{keyword1; keyword2; keyword3.}%WSPC
\end{abstract}

%\ccode{PACS Nos.: nn.nn.Xx; nn.nn.XX}%WSPC

%REVTEX4
%REVTEX4

%\begin{keyword}%ELSART
%  keyword 1\sep keyword 2\sep keyword 3
% \sep keyword 4\sep keyword 5\sep keyword 6
% \PACS nn.nn.Xx\sep nn.nn.Xx\sep nn.nn.Xn\sep nn.nn.Xx
%\end{keyword}%ELSART

%\pacs{nn.nn.Xx, nn.nn.Xx, nn.nn.Xx, nn.nn.Xx}%IOPART
%\submitto{Journal}

\preprint{TH-1479}%REVTEX4

\maketitle

\section{Introduction}
\label{sec:intro}

\noindent
Recently, supersymmetry (SUSY) and shape invariance in quantum mechanical
matrix models have attracted much attention in the literature, e.g.,
Refs.~\cite{ACD88,Fu93,HGB95,ACIN97,LRSFV98,DC99,LC99,Tr99,Tr00,LRBV01,%
IN03,IKNN06,FMN10,NK11} in various physical contexts. To avoid confusion, we
note that we here mean SUSY between two matrix Schr\"{o}dinger operators
intertwined by a matrix linear differential operator but \emph{not} SUSY between
two scalar Schr\"{o}dinger operators in a matrix superHamiltonian intertwined by
a scalar linear differential operator.

On the other hand, the framework of $\cN$-fold SUSY~\cite{AIS93,AST01b,AS03}
has been shown to be quite fruitful among several generalizations of ordinary SUSY
especially since the establishment of its equivalence with weak quasi-solvability
in Ref.~\cite{AST01b}, for a review see Ref.~\cite{Ta09}. Due to the facts that
the $\cN=1$ case corresponds to ordinary SUSY and that shape invariance
automatically implies weak quasi-solvability, $\cN$-fold SUSY contains both
ordinary SUSY and shape invariance as its particular cases.

Hence, it is quite natural to ask whether a formulation of $\cN$-fold SUSY is
possible for matrix systems. To the best of our knowledge, there was so far only
one such an attempt which corresponds to the $\cN=2$ case for $2\times 2$
matrix models~\cite{ACIN97}. However, the analysis there was quite 
restrictive, was devoted mostly to the particular cases where the systems were
available by two successive SUSY transformations, and resorted to the involved
assumptions and ansatz. As a consequence, its formulation does not have a form
which is suitable for discussing general aspects, especially those which were
established and appreciated later after that work.

In this article, we formulate for the first time $\cN$-fold SUSY for a system
composed of matrix Schr\"{o}dinger operators  for all positive integral $\cN$
in such a general fashion that the recent crucial developments in the field
are fully incorporated in the formalism. To see such a system actually exists,
we construct as an illustration general $2\times 2$ Hermitian matrix $2$-fold
SUSY systems without any assumption or ansatz. We find that there are two
inequivalent systems, both of which are characterized by two arbitrary
real scalar functions. Intriguingly, one of the systems does not
admit reduction to the most general $2$-fold SUSY scalar system.

We organize this article as follows. In the next section, we generically define
$\cN$-fold SUSY in quantum mechanical matrix models. Then, we investigate
in detail $2\times 2$ Hermitian matrix systems for $\cN=2$ in Section~\ref{sec:2f}.
We explicitly solve all the conditions for $2$-fold SUSY to obtain general form of
the latter systems. In the last section, we refer to several future issues to be
followed after this work.

\section{General Setting}
%\label{sec:gen}
%\nosections%IOPART
A quantum mechanical system we shall consider here is a pair of $n\times n$
matrix Schr\"{o}dinger operators
\begin{align}
H^{\pm}=-\frac{1}{2}I_{n}\frac{\rmd^{2}}{\rmd q^{2}}+V^{\pm}(q),
\label{eq:H+-}
\end{align}
where $I_{n}$ is an $n\times n$ unit matrix and each $V^{\pm}(q)$ is an $n\times n$
matrix-valued complex function. Let us introduce a pair of $n\times n$ matrix linear
differential operators of order $\cN$
\begin{subequations}
\label{eqs:PN+-}
\begin{align}
P_{\cN}^{-}&=I_{n}\frac{\rmd^{\cN}}{\rmd q^{\cN}}+\sum_{k=0}^{\cN-1}w_{k}(q)
 \frac{\rmd^{k}}{\rmd q^{k}},\\
P_{\cN}^{+}&=(-1)^{\cN}I_{n}\frac{\rmd^{\cN}}{\rmd q^{\cN}}+\sum_{k=0}^{\cN-1}
 (-1)^{k}\frac{\rmd^{k}}{\rmd q^{k}}w_{k}(q),
\end{align}
\end{subequations}
where $w_{k}(q)$ ($k=0,\dots,\cN-1$) are $n\times n$ matrix-valued complex functions.
Then, the system (\ref{eq:H+-}) is said to be \emph{$\cN$-fold supersymmetric}
with respect to (\ref{eqs:PN+-}) if the following relations are all satisfied:
\begin{subequations}
\label{eqs:NS}
\begin{align}
&P_{\cN}^{\mp}H^{\mp}-H^{\pm}P_{\cN}^{\mp}=0,
\label{eq:NS1}\\
&P_{\cN}^{\mp}P_{\cN}^{\pm}=2^{\cN}\left[(H^{\pm}+C_{0})^{\cN}+\sum_{k=1}^{\cN-1}C_{k}
 (H^{\pm}+C_{0})^{\cN-k-1}\right],
\label{eq:NS2}
\end{align}
\end{subequations}
where $C_{k}$ ($k=0,\dots,\cN-1$) are $n\times n$ constant matrices. It is evident
that in the case of $n=1$ the above definition of $\cN$-fold SUSY reduces to
the ordinary one for a pair of scalar Schr\"{o}dinger operators. As in the scalar case,
the first relation (\ref{eq:NS1}) immediately implies almost isospectrality of
$H^{\pm}$ and \emph{weak quasi-solvability}
$H^{\pm}\ker P_{\cN}^{\pm}\subset\ker P_{\cN}^{\pm}$.
We note that in contrast to the formulation in Ref.~\cite{ACIN97} where its focus
was on hidden symmetry operators characterized as the deviation from the second
algebraic relation (\ref{eq:NS2}) we have treated it as a part of the definition of
$\cN$-fold SUSY. See also the discussion in Section~\ref{sec:discus}, item~1 for
its relevance.

\section{Two-by-two Hermitian $2$-fold Supersymmetry}
\label{sec:2f}

As an example, let us construct the most general $2\times 2$ Hermitian matrix
$2$-fold SUSY systems which are defined in the Hilbert space $L^{2}$ of
two-component functions equipped with the inner product defined by
\begin{align}
(\phi,\psi)=\int_{S}\rmd q\,\phi^{\dagger}(q)\psi(q),\quad\phi,\psi\in L^{2}(S),\quad
 S\subset\bbR,
\label{eq:inner}
\end{align}
where the Hermitian conjugate $\dagger$ is as usual the combination of complex
conjugate and transposition. The most general form of a pair of $2\times 2$
Hermitian matrix Schr\"{o}dinger operators is given by
\begin{align}
H^{\pm}=-\frac{1}{2}I_{2}\frac{\rmd^{2}}{\rmd q^{2}}
 +\sum_{\mu=0}^{3}V_{\mu}^{\pm}(q)\sigma_{\mu},
\label{eq:H+-2}
\end{align}
where $\sigma_{0}=I_{2}$ is a $2\times 2$ unit matrix, $\sigma_{i}$ ($i=1,2,3$) are
the Pauli matrices, and $V_{\mu}(q)$ ($\mu=0,\dots,3$) are all real scalar functions.
Components of $2\times 2$ matrix $2$-fold supercharges have the following form
\begin{subequations}
\label{eqs:P2+-}
\begin{align}
P_{2}^{-}&=I_{2}\frac{\rmd^{2}}{\rmd q^{2}}+\left(\sum_{\mu=0}^{3}w_{1\mu}(q)
 \sigma_{\mu}\right)\frac{\rmd}{\rmd q}+\sum_{\mu=0}^{3}w_{0\mu}(q)\sigma_{\mu},\\
P_{2}^{+}&=I_{2}\frac{\rmd^{2}}{\rmd q^{2}}-\frac{\rmd}{\rmd q}\left(\sum_{\mu=0}^{3}
 w_{1\mu}(q)\sigma_{\mu}\right)+\sum_{\mu=0}^{3}w_{0\mu}(q)\sigma_{\mu},
\end{align}
\end{subequations}
where $w_{1\mu}(q)$ and $w_{0\mu}(q)$ ($\mu=0,\dots,3$) are all real scalar
functions. To investigate the $\cN$-fold SUSY condition (\ref{eqs:NS}) for the
$\cN=2$ case under consideration, we first note that $P_{2}^{-}H^{-}-H^{+}P_{2}^{-}=0$
implies $P_{2}^{+}H^{+}-H^{-}P_{2}^{+}=0$ and vice versa since they are Hermitian
conjugate with each other with respect to the inner product (\ref{eq:inner}), thanks
to the choices (\ref{eq:H+-2}) and (\ref{eqs:P2+-}). Hence, it is sufficient to
study only the former. A direct calculation shows that the intertwining relation
$P_{2}^{-}H^{-}-H^{+}P_{2}^{-}=0$ holds if and only if the following set of conditions
is satisfied:
\begin{align}
&V_{\mu}^{+}-V_{\mu}^{-}=w'_{1\mu},
\label{eq:co1}\\
&w''_{10}+2w'_{00}+4V_{0}^{-\prime}-2\sum_{\mu=0}^{3}w_{1\mu}(V_{\mu}^{+}
 -V_{\mu}^{-})=0,
\label{eq:co2}\\
&w''_{1i}+2w'_{0i}+4V_{i}^{-\prime}-2w_{1i}(V_{0}^{+}-V_{0}^{-})-2w_{10}(V_{i}^{+}
 -V_{i}^{-})=0,
\label{eq:co3}\\
&\sum_{j,k=1}^{3}\epsilon_{ijk}w_{1j}(V_{k}^{+}+V_{k}^{-})=0,
\label{eq:co4}\\
&w''_{00}+2V_{0}^{-\prime\prime}+2\sum_{\mu=0}^{3}\left[w_{1\mu}V_{\mu}^{-\prime}
 -w_{0\mu}(V_{\mu}^{+}-V_{\mu}^{-})\right]=0,
\label{eq:co5}\\
&w''_{0i}+2V_{i}^{-\prime\prime}+2w_{1i}V_{0}^{-\prime}+2w_{10}V_{i}^{-\prime}
 -2w_{0i}(V_{0}^{+}-V_{0}^{-})-2w_{00}(V_{i}^{+}-V_{i}^{-})=0,
\label{eq:co6}\\
&\sum_{j,k=1}^{3}\epsilon_{ijk}\left[w_{1j}V_{k}^{-\prime}
 +w_{0j}(V_{k}^{+}+V_{k}^{-})\right]=0.
\label{eq:co7}
\end{align}
On the other hand, we find that the $2$-fold superalgebra $P_{2}^{\mp}P_{2}^{\pm}
=4\left[(H^{\pm}+C_{0})^{2}+C_{1}\right]$ with Hermitian constant matrices $C_{k}=
\sum_{\mu=0}^{3}C_{k\mu}\sigma_{\mu}$ ($C_{k\mu}\in\bbR$, $k=0,1$) holds for
the upper sign if and only if
\begin{align}
&4V_{0}^{+}=3w'_{10}-2w_{00}+\sum_{\mu=0}^{3}(w_{1\mu})^{2}-4C_{00},
\label{eq:c+1}\\
&4V_{i}^{+}=3w'_{1i}-2w_{0i}+2w_{10}w_{1i}-4C_{0i},
\label{eq:c+2}\\
&\sum_{j,k=1}^{3}\epsilon_{ijk}w_{1j}(w'_{1k}-w_{0k})=0,
\label{eq:c+3}\\
&2V_{0}^{+\prime\prime}-4\sum_{\mu=0}^{3}(V_{\mu}^{+}+C_{0\mu})^{2}-4C_{10}
 =&\notag\\
&\hspace*{30pt}w'''_{10}-w''_{00}+\sum_{\mu=0}^{3}\left[w_{1\mu}w''_{1\mu}
 +w'_{1\mu}w_{0\mu}-w_{1\mu}w'_{0\mu}-(w_{0\mu})^{2}\right],
\label{eq:c+4}\\[5pt]
&2V_{i}^{+\prime\prime}-8(V_{0}^{+}+C_{00})(V_{i}^{+}+C_{0i})-4C_{1i}=
 w'''_{1i}-w''_{0i}\notag\\
&\hspace*{30pt}+w''_{10}w_{1i}+w_{10}w''_{1i}-w'_{00}w_{1i}+w_{00}w'_{1i}+w'_{10}w_{0i}
 -w_{10}w'_{0i}-2w_{00}w_{0i},
\label{eq:c+5}\\
&\sum_{j,k=1}^{3}\epsilon_{ijk}(w''_{1j}w_{1k}+w'_{1j}w_{0k}+w_{1j}w'_{0k})=0,
\label{eq:c+6}
\end{align}
and for the lower sign if and only if
\begin{align}
&4V_{0}^{-}=-w'_{10}-2w_{00}+\sum_{\mu=0}^{3}(w_{1\mu})^{2}-4C_{00},
\label{eq:c-1}\\
&4V_{i}^{-}=-w'_{1i}-2w_{0i}+2w_{10}w_{1i}-4C_{0i},
\label{eq:c-2}\\
&\sum_{j,k=1}^{3}\epsilon_{ijk}w_{1j}w_{0k}=0,
\label{eq:c-3}\\
&2V_{0}^{-\prime\prime}-4\sum_{\mu=0}^{3}(V_{\mu}^{-}+C_{0\mu})^{2}-4C_{10}=
 -w''_{00}+\sum_{\mu=0}^{3}\left[w'_{1\mu}w_{0\mu}+w_{1\mu}w'_{0\mu}
 -(w_{0\mu})^{2}\right],
\label{eq:c-4}\\
&2V_{i}^{-\prime\prime}-8(V_{0}^{-}+C_{00})(V_{i}^{-}+C_{0i})-4C_{1i}=\notag\\
&\hspace*{30pt}-w''_{0i}+w'_{00}w_{1i}+w_{00}w'_{1i}+w'_{10}w_{0i}+w_{10}w'_{0i}
 -2w_{00}w_{0i},
\label{eq:c-5}\\
&\sum_{j,k=1}^{3}\epsilon_{ijk}(w'_{1j}w_{0k}+w_{1j}w'_{0k})=0.
\label{eq:c-6}
\end{align}
The formulas (\ref{eq:c+1}), (\ref{eq:c+2}), (\ref{eq:c-1}), and (\ref{eq:c-2}) determine
the form of the potentials $V_{\mu}^{\pm}$ and are compatible with the conditions
(\ref{eq:co1}), (\ref{eq:co2}), and (\ref{eq:co3}). Then, the conditions (\ref{eq:co4}),
(\ref{eq:c+3}), and (\ref{eq:c-3}) are identical with
\begin{align}
\sum_{j,k=1}^{3}\epsilon_{ijk}w_{1j}w_{0k}=\sum_{j,k=1}^{3}\epsilon_{ijk}w_{1j}w'_{1k}=
 \sum_{j,k=1}^{3}\epsilon_{ijk}w_{1j}C_{0k}=0.
\end{align}
The most general solutions to the latter set of conditions are given by
\begin{align}
w_{1i}=C_{0i}v_{1},\qquad w_{0i}=C_{0i}v_{0},
\label{eq:v1v0}
\end{align}
where $v_{1}$ and $v_{0}$ are at present arbitrary scalar functions. Substituting
(\ref{eq:v1v0}) into (\ref{eq:c+1}), (\ref{eq:c+2}), (\ref{eq:c-1}), and (\ref{eq:c-2}),
we obtain
\begin{subequations}
\label{eqs:V1}
\begin{align}
4V_{0}^{+}&=3w'_{10}-2w_{00}+(w_{10})^{2}+C^{2}(v_{1})^{2}-4C_{00},\\
4V_{i}^{+}&=C_{0i}(3v'_{1}-2v_{0}+2w_{10}v_{1}-4),\\
4V_{0}^{-}&=-w'_{10}-2w_{00}+(w_{10})^{2}+C^{2}(v_{1})^{2}-4C_{00},\\
4V_{i}^{-}&=C_{0i}(-v'_{1}-2v_{0}+2w_{10}v_{1}-4),
\end{align}
\end{subequations}
where $C^{2}=\sum_{i=1}^{3}(C_{0i})^{2}$. The solutions (\ref{eq:v1v0}) automatically
satisfy (\ref{eq:co7}), (\ref{eq:c+6}), and (\ref{eq:c-6}). With the substitution of
(\ref{eq:v1v0}) and (\ref{eqs:V1}) into the remaining conditions,
(\ref{eq:co5}) and (\ref{eq:co6}) are respectively identical with
\begin{align}
&w'''_{10}-w_{10}w''_{10}-2(w'_{10})^{2}+4w'_{10}w_{00}+2w_{10}w'_{00}-2(w_{10})^{2}
 w'_{10}\notag\\
&-C^{2}\left[v_{1}v''_{1}+2(v'_{1})^{2}-4v'_{1}v_{0}-2v_{1}v'_{0}+2w'_{10}(v_{1})^{2}
 +4w_{10}v_{1}v'_{1}\right]=0,
\label{eq:co5'}\\
&v'''_{1}-w''_{10}v_{1}-4w'_{10}v'_{1}-w_{10}v''_{1}+4w'_{10}v_{0}+2w_{10}v'_{0}
 +2w'_{00}v_{1}+4w_{00}v'_{1}\notag\\
&-4w_{10}w'_{10}v_{1}-2(w_{10})^{2}v'_{1}-2C^{2}(v_{1})^{2}v'_{1}=0,
\label{eq:co6'}
\end{align}
(\ref{eq:c+4}) and (\ref{eq:c+5}) are respectively with
\begin{align}
&2w'''_{10}-5(w'_{10})^{2}+8w'_{10}w_{00}+4w_{10}w'_{00}-6(w_{10})^{2}w'_{10}
 +4(w_{10})^{2}w_{00}-(w_{10})^{4}\notag\\
&-C^{2}\bigl[5(v'_{1})^{2}-8v'_{1}v_{0}-4v_{1}v'_{0}
 +6w'_{10}(v_{1})^{2}+12w_{10}v_{1}v'_{1}-8w_{10}v_{1}v_{0}\notag\\
&-4w_{00}(v_{1})^{2}+6(w_{10})^{2}(v_{1})^{2}\bigr]-C^{4}(v_{1})^{4}-16C_{10}=0,
\label{eq:c+4'}\\
&C_{0i}\bigl[v'''_{1}-5w'_{10}v'_{1}+4w'_{10}v_{0}+2w_{10}v'_{0}+2w'_{00}v_{1}
 +4w_{00}v'_{1}-6w_{10}w'_{10}v_{1}-3(w_{10})^{2}v'_{1}\notag\\
&+2(w_{10})^{2}v_{0}+4w_{10}w_{00}v_{1}-2(w_{10})^{3}v_{1}
 -C^{2}(v_{1})^{2}(3v'_{1}-2v_{0}+2w_{10}v_{1})\bigr]-8C_{1i}=0,
\label{eq:c+5'}
\end{align}
and (\ref{eq:c-4}) and (\ref{eq:c-5}) are respectively with
\begin{align}
&2w'''_{10}-4w_{10}w''_{10}-3(w'_{10})^{2}+8w'_{10}w_{00}+4w_{10}w'_{00}
 -2(w_{10})^{2}w'_{10}-4(w_{10})^{2}w_{00}\notag\\
&+(w_{10})^{4}-C^{2}\bigl[4v_{1}v''_{1}+3(v'_{1})^{2}-8v'_{1}v_{0}-4v_{1}v'_{0}
 +2w'_{10}(v_{1})^{2}+4w_{10}v_{1}v'_{1}\notag\\
&+8w_{10}v_{1}v_{0}+4w_{00}(v_{1})^{2}-6(w_{10})^{2}(v_{1})^{2}\bigr]
 +C^{4}(v_{1})^{4}+16C_{10}=0,
\label{eq:c-4'}\\
&C_{0i}\bigl[v'''_{1}-2w''_{10}v_{1}-3w'_{10}v'_{1}-2w_{10}v''_{1}+4w'_{10}v_{0}
 +2w_{10}v'_{0}+2w'_{00}v_{1}+4w_{00}v'_{1}\notag\\
&-2w_{10}w'_{10}v_{1}-(w_{10})^{2}v'_{1}-2(w_{10})^{2}
 v_{0}-4w_{10}w_{00}v_{1}+2(w_{10})^{3}v_{1}\notag\\
&+C^{2}(v_{1})^{2}(-v'_{1}-2v_{0}+2w_{10}v_{1})\bigr]+8C_{1i}=0.
\label{eq:c-5'}
\end{align}
It is evident that the equations (\ref{eq:c+5'}) and (\ref{eq:c-5'}) have the
trivial solutions $C_{1i}=C_{0i}=0$. On the other hand, for non-trivial solutions
the combination $C_{1i}/C_{0i}:=\tilde{C}$ ($i=1,2,3$) should not depend on
the index $i$.

Let us first consider the set of conditions (\ref{eq:co5'}) and (\ref{eq:co6'}).
We find that the two combinations $w_{10}\times$(\ref{eq:co5'})$+C^{2}v_{1}\times
$(\ref{eq:co6'}) and $v_{1}\times$(\ref{eq:co5'})$+w_{10}\times$(\ref{eq:co6'})
are total differentials and thus are integrated respectively as
\begin{align}
&2w_{10}w''_{10}-(w'_{10})^{2}-2(w_{10})^{2}w'_{10}+4(w_{10})^{2}w_{00}-(w_{10})^{4}
 \notag\\
&+C^{2}\bigl[2v_{1}v''_{1}-(v'_{1})^{2}-2w'_{10}(v_{1})^{2}-4w_{10}v_{1}v'_{1}
 +8w_{10}v_{1}v_{0}\notag\\
&+4w_{00}(v_{1})^{2}-6(w_{10})^{2}(v_{1})^{2}\bigr]-C^{4}(v_{1})^{4}=D_{1},
\label{eq:d1}
\end{align}
and
\begin{align}
&w''_{10}v_{1}-w'_{10}v'_{1}+w_{10}v''_{1}-2w_{10}w'_{10}v_{1}-(w_{10})^{2}v'_{1}
 +2(w_{10})^{2}v_{0}+4w_{10}w_{00}v_{1}\notag\\
&-2(w_{10})^{3}v_{1}-C^{2}(v_{1})^{2}(v'_{1}-2v_{0}+2w_{10}v_{1})=D_{2},
\label{eq:d2}
\end{align}
where $D_{1}$ and $D_{2}$ are integral constants. It is easily checked that
(\ref{eq:d1}) and (\ref{eq:d2}) are compatible with all the remaining conditions
(\ref{eq:c+4'})--(\ref{eq:c-5'}) if and only if
\begin{align}
D_{1}=16C_{10},\qquad D_{2}=8C_{1i}/C_{0i}=8\tilde{C}.
\end{align}
Hence, the only remaining problem is to analyze (\ref{eq:d1}) and (\ref{eq:d2}).
They can be regarded as simultaneous linear equations for $w_{00}$ and $v_{0}$.
For the non-degenerate case $v_{1}\neq C^{-1}w_{10}$, they are uniquely
solved as
\begin{align}
&4\left[(w_{10})^{2}-C^{2}(v_{1})^{2}\right]^{2}w_{00}=(w_{10})^{2}\bigl[-2w_{10}
 w''_{10}+(w'_{10})^{2}+2(w_{10})^{2}w'_{10}+(w_{10})^{4}+16C_{10}\bigr]\notag\\
&+C^{2}\bigl[2w_{10}w''_{10}(v_{1})^{2}+(w'_{10})^{2}(v_{1})^{2}-4w_{10}w'_{10}v_{1}v'_{1}
 +2(w_{10})^{2}v_{1}v''_{1}+(w_{10})^{2}(v'_{1})^{2}\notag\\
&-4(w_{10})^{2}w'_{10}(v_{1})^{2}
 -(w_{10})^{4}(v_{1})^{2}-32\tilde{C}w_{10}v_{1}+16C_{10}(v_{1})^{2}\bigr]\notag\\
&-C^{4}(v_{1})^{2}\bigl[2v_{1}v''_{1}-(v'_{1})^{2}-2w'_{10}(v_{1})^{2}+(w_{10})^{2}
 (v_{1})^{2}\bigr]+C^{6}(v_{1})^{6},
\label{eq:w00}
\end{align}
and
\begin{align}
&2\left[(w_{10})^{2}-C^{2}(v_{1})^{2}\right]^{2}v_{0}=w_{10}\bigl[w_{10}w''_{10}v_{1}
 -(w'_{10})^{2}v_{1}+w_{10}w'_{10}v'_{1}-(w_{10})^{2}v''_{1}+(w_{10})^{3}v'_{1}\notag\\
&+(w_{10})^{4}v_{1}+8\tilde{C}w_{10}-16C_{10}v_{1}\bigr]+C^{2}v_{1}\bigl[
 -w''_{10}(v_{1})^{2}+w'_{10}v_{1}v'_{1}+2w_{10}v_{1}v''_{1}-w_{10}(v'_{1})^{2}\notag\\
&-2(w_{10})^{2}v_{1}v'_{1}-2(w_{10})^{3}(v_{1})^{2}+8\tilde{C}v_{1}\bigr]
 +C^{4}(v_{1})^{4}(v'_{1}+w_{10}v_{1}).
\label{eq:v0}
\end{align}
Finally, the general form of a $2\times 2$ Hermitian matrix $2$-fold SUSY system
for the non-degenerate case is given by
\begin{align}
H^{+}=&\;-\frac{1}{2}\frac{\rmd^{2}}{\rmd q^{2}}+\frac{1}{4}\left[3w'_{10}-2w_{00}
 +(w_{10})^{2}+C^{2}(v_{1})^{2}\right]\notag\\
&\;+\frac{1}{4}(3v'_{1}-2v_{0}+2w_{10}v_{1})\sum_{i=1}^{3}C_{0i}\sigma_{i}-C_{0},
\label{eq:ndH+}\\
H^{-}=&\;-\frac{1}{2}\frac{\rmd^{2}}{\rmd q^{2}}+\frac{1}{4}\left[-w'_{10}-2w_{00}
 +(w_{10})^{2}+C^{2}(v_{1})^{2}\right]\notag\\
&\;+\frac{1}{4}(-v'_{1}-2v_{0}+2w_{10}v_{1})\sum_{i=1}^{3}C_{0i}\sigma_{i}-C_{0},\\
P_{2}^{-}=&\;\frac{\rmd^{2}}{\rmd q^{2}}+\left(w_{10}+v_{1}\sum_{i=1}^{3}C_{0i}
 \sigma_{i}\right)\frac{\rmd}{\rmd q}+w_{00}+v_{0}\sum_{i=1}^{3}C_{0i}\sigma_{i}.
\label{eq:ndP2}
\end{align}
The two functions $w_{00}$ and $v_{0}$ in the above can be eliminated by using
(\ref{eq:w00}) and (\ref{eq:v0}). Hence, the system can be expressed solely in
terms of the two functions $w_{10}$ and $v_{1}$.

For the degenerate case $v_{1}=C^{-1}w_{10}$, the two equations (\ref{eq:d1})
and (\ref{eq:d2}) are not independent and are equivalent with the following
single equation
\begin{align}
4(w_{10})^{2}(w_{00}+Cv_{0})=-2w_{10}w''_{10}+(w'_{10})^{2}+4(w_{10})^{2}w'_{10}
 +4(w_{10})^{4}+8C_{10},
\end{align}
with $C_{10}=\tilde{C}C$. Hence, we can again eliminate two of the four functions,
e.g., $v_{1}$ and $v_{0}$. The general form of a $2\times 2$ Hermitian matrix $2$-fold
SUSY system for the degenerate case is given by
\begin{align}
H^{+}=&\;-\frac{1}{2}\frac{\rmd^{2}}{\rmd q^{2}}+\frac{1}{4}\left[3w'_{10}-2w_{00}
 +2(w_{10})^{2}\right]\notag\\
&\;+\frac{1}{4C}\left[w'_{10}+2w_{00}+\frac{w''_{10}}{w_{10}}-\frac{(w'_{10})^{2}}{
 2(w_{10})^{2}}-\frac{4\tilde{C}C}{(w_{10})^{2}}\right]\sum_{i=1}^{3}C_{0i}\sigma_{i}
 -C_{0},
\label{eq:dH+}\\
H^{-}=&\;-\frac{1}{2}\frac{\rmd^{2}}{\rmd q^{2}}+\frac{1}{4}\left[-w'_{10}-2w_{00}
 +2(w_{10})^{2}\right]\notag\\
&\;+\frac{1}{4C}\left[-3w'_{10}+2w_{00}+\frac{w''_{10}}{w_{10}}-\frac{(w'_{10})^{2}}{
 2(w_{10})^{2}}-\frac{4\tilde{C}C}{(w_{10})^{2}}\right]\sum_{i=1}^{3}C_{0i}\sigma_{i}
 -C_{0},\\
P_{2}^{-}=&\;\frac{\rmd^{2}}{\rmd q^{2}}+w_{10}\left(1+C^{-1}\sum_{i=1}^{3}C_{0i}
 \sigma_{i}\right)\frac{\rmd}{\rmd q}\notag\\
&+w_{00}+\frac{1}{C}\left[w'_{10}-w_{00}+(w_{10})^{2}-\frac{w''_{10}}{2w_{10}}
 +\frac{(w'_{10})^{2}}{4(w_{10})^{2}}+\frac{2\tilde{C}C}{(w_{10})^{2}}\right]
 \sum_{i=1}^{3}C_{0i}\sigma_{i}.
\label{eq:dP2}
\end{align}
It is interesting to note that in the limit $C_{0i}\to 0$ ($i=1,2,3$), the non-degenerate
system (\ref{eq:ndH+})--(\ref{eq:ndP2}) reduces to the most general $2$-fold SUSY
scalar system~\cite{AST01b,AICD95,AIN95b} while the degenerate system
(\ref{eq:dH+})--(\ref{eq:dP2}) does \emph{not}.

\section{Discussion and Summary}
\label{sec:discus}

In this article, we have for the first time formulated generically $\cN$-fold SUSY
in quantum mechanical matrix models and constructed the general $2\times 2$
Hermitian matrix $2$-fold SUSY systems without recourse to any assumption
or ansatz. In addition to the detailed studies for larger $n\times n$ matrices
($n>2$) and $\cN>2$ cases, there are many future issues to be followed after
this work as the following:\\

\noindent
1. First of all, it is important to clarify general aspects of $\cN$-fold SUSY in matrix
 systems, as were done in~\cite{AST01b,AS03} for scalar systems. In the scalar case,
 there are two significant features, namely, the equivalence between $\cN$-fold SUSY
 and weak quasi-solvability and the equivalence between the conditions (\ref{eq:NS1})
 and (\ref{eq:NS2}). In the case of $2\times 2$ Hermitian matrix systems,
 however, it does not seem that the conditions (\ref{eq:co4}) and (\ref{eq:co7}) coming
 from the former are equivalent with the conditions (\ref{eq:c+3}), (\ref{eq:c+6}),
 (\ref{eq:c-3}), and (\ref{eq:c-6}) coming from the latter although the other conditions
 are certainly equivalent. That is exactly the reason why we considered the both to
 derive the formula (\ref{eq:v1v0}). We expect that the general approach~\cite{Ta11a}
 for the  scalar case recently proposed by us would be also efficient for the matrix
 case.\\

\noindent
2. In the scalar case, the systematic algorithm for constructing an $\cN$-fold SUSY
 system~\cite{GT05} based on quasi-solvability has shown to be quite effective.
 Hence, its generalization to the matrix case is desirable. It would also enable us to
 connect directly the possible types of matrix $\cN$-fold SUSY systems with the
 possible linear spaces of multi-component functions preserved by a second-order
 matrix linear differential operator. For example, it is interesting to see the connection
 with the quasi-solvable matrix operators constructed from the generators of
 $\mathfrak{sl}(2)$ in Ref.~\cite{Zh97}.\\

\noindent
3. Shape invariance is a well-known sufficient condition for \emph{solvability}~\cite{Ge83}.
 It means in particular that it always implies $\cN$-fold SUSY in the scalar case.
 In fact, some shape-invariant scalar potentials were systematically constructed
 as particular cases of $\cN$-fold SUSY with intermediate Hamiltonians~\cite{BT09,BT10}.
 Recently, several shape-invariant matrix potentials were constructed in
 Refs.~\cite{Fu93,DC99,IKNN06,FMN10,NK11}, and we expect that our formulation
 of $\cN$-fold SUSY would be also quite efficient in constructing shape-invariant
 matrix models.\\

\noindent
4. Extension to more general second-order matrix linear differential operators would
 be possible, e.g., by admitting a non-diagonal second-order operator and by adding
 a matrix-valued first-order operator. In particular, a quantum mechanical matrix
 model with matrix-valued position-dependent mass would be an interesting candidate
 as a natural generalization of $\cN$-fold SUSY in scalar quantum systems with
 position-dependent mass~\cite{Ta06a}.\\

\noindent
5. In the scalar case, there are several intimate relations between $\cN$-fold SUSY
 and $\cN$th-order paraSUSY~\cite{BT09,BT10,Ta07a,Ta07c}. We expect that we can
 formulate higher-order paraSUSY in quantum mechanical matrix models in a way
 such that the relations to $\cN$-fold SUSY in the scalar case remain intact in
 a matrix case. Extension of higher-order $\cN$-fold paraSUSY~\cite{Ta07b} to
 matrix systems would be also possible.

%\section*{Acknowledgments}
%\begin{acknowledgments}%REVTEX4
%\ack%IOPART

%\end{acknowledgments}%REVTEX4

%\appendix

%\section*{References}%IOPART

\bibliography{refsels}%BIB-FILE
\bibliographystyle{npb}%BST-FILE
%\begin{thebibliography}{99}
%\def\J#1#2#3#4{{\sl #1} {\bf #2} (#3) #4}

%\bibitem{}
%Author 1, Author 2 and Author 3,
%\J{Journal}{Volume}{Year}{Page}.

%\end{thebibliography}

\end{document}